\documentclass[twocolumn,fleqn,natbib]{svjour2}
\usepackage{amsmath}
\usepackage{citesort}
\bibpunct{[}{]}{;}{n}{}{,} 
\smartqed  
\usepackage{graphicx}
\usepackage{mathptmx}      
\journalname{Granular Matter}
\begin{document}

\title{Coefficient of restitution and linear dashpot model revisited\thanks{This research was supported by German Science Foundation (Grant PO 472/15-1).}}


\titlerunning{Coefficient of restitution and linear dashpot model revisited} 

\author{Thomas Schwager \and Thorsten P\"{o}schel}

\authorrunning{T. Schwager and T. P\"oschel}

\institute{
Thomas Schwager \at
Charit\'e\\ 
Augustenburger Platz\\ 
10439 Berlin\\ 
Germany\\
\email{thomas.schwager@charite.de}
\and
Thorsten P\"{o}schel \at
Charit\'e\\ 
Augustenburger Platz\\ 
10439 Berlin\\ 
Germany\\
\email{thorsten.poeschel@charite.de}
}

\date{Received: \today }

\maketitle

\begin{abstract}
  With the assumption of a linear-dashpot interaction force, we compute the
  coefficient of restitution as a function of the elastic and dissipative
  material constants, $k$ and $\gamma$ by integrating Newton's equation of
  motion for an isolated pair of colliding particles.  It turns out that an
  expression $\varepsilon(k,\gamma)$ widely used in the literature is even
  qualitatively incorrect due to an incorrect definition of the duration of the
  collision.  
\end{abstract}

\section{Introduction}
\label{intro}

The (normal) coefficient of restitution of colliding spherical particles
relates the post-collisional velocities $\vec{v}_i^\prime$ and
$\vec{v}_j^\prime$ and the corresponding pre-collisional velocities
$\vec{v}_i$ and $\vec{v}_j$,
\begin{equation}
  \label{eq:EMD}
  \begin{split}
    \vec{v}_i^{\,\prime} = & \vec{v}_i - \frac{1+\varepsilon}{2}\left[\left(\vec{v}_i-\vec{v}_j\right)\cdot \vec{e}_{ij}\right]\vec{e}_{ij}\,\\
    \vec{v}_j^{\,\prime} = & \vec{v}_j + \frac{1+\varepsilon}{2}\left[\left(\vec{v}_i-\vec{v}_j\right)\cdot
    \vec{e}_{ij}\right]\vec{e}_{ij}\,
  \end{split}
\end{equation}
where 
\begin{equation}
\vec{e}_{ij}\equiv
\frac{\vec{r}_i-\vec{r}_j}{\left|\vec{r}_i-\vec{r}_j\right|}
\end{equation}
is the unit vector at the instance of the collision. For simplicity,
we consider particles of identical mass - the generalization to unequal masses
is straightforward.

The coefficient of restitution is the most fundamental quantity in the theory
of dilute granular systems. This coefficient together with the assumption of
\textit{molecular chaos} is the foundation of the Kinetic Theory of granular
gases, based on the Boltzmann or Enskog kinetic equation, see, e.g.,
\cite{Brilliantov,PoeschelBrilliantov:2003,PoeschelLuding:2001} and many
references therein. On the other hand, the coefficient of restitution is also
essential for event-driven Molecular Dynamics simulations which allow for a
significant speed-up as compared to traditional force-based simulations, see,
e.g., \cite{Algo} for a detailed discussion. The most important precondition
of Eq. \eqref{eq:EMD} is the assumption of exclusively binary collision, that
is, the system is assumed to be dilute enough such that multiple-particle
contacts can be neglected. The latter condition seems to be rather restrictive,
however, it was shown in many simulations that event-driven simulations are
applicable up to rather high density \cite{MeersonPoeschelBromberg:2003}.

The coefficient of restitution is, however, not a fundamental material or
particle property. Instead, the particle interaction forces and Newton's
equation of motion govern the dynamics of a mechanical many-particle system.
Therefore, if we wish to use the concept of the coefficient of restitution and
the simple equation \eqref{eq:EMD} to compute the dynamics of a granular
system, we have to assure that Eq. \eqref{eq:EMD} yields the same
post-collisional velocities as Newton's equation of motion would do.

\section{Coefficient of restitution and Newton's equation of motion for
  frictionless spherical particles}

Let us consider the relation between the coefficient of restitution and the
solution of Newton's equation of motion. Assume two approaching particles at
velocities $\vec{v}_1$ and $\vec{v}_2$ come
into contact at time $t=0$. From this instant on they deform one another until
the contact is lost at time $t=t_c$. We shall consider the end of the collision
at time $t_c$ separately below. For sufficiently short collisions we can treat the unity vector $\vec{e}$ as a constant. Then we can describe the collision by the mutual deformation
\begin{equation}
\xi(t)\equiv\max\left(0,~2R-\left|\vec{r}_1-\vec{r}_2\right|\right)\,
\end{equation}
where $\vec{r}_1(t)$ and $\vec{r}_2(t)$ are the time-dependent positions of
the particles and $R$ is their radius. The interaction force between the
particles is model specific, e.g. \cite{Algo,SchaeferDippelWolf:1995}, and, in
general, a function of the material parameters, the particle masses and radii,
the deformation $\xi$ and the deformation rate $\dot{\xi}$. Its time
dependence is expressed via $F=F\left(\xi(t),\dot{\xi}(t)\right)$. Newton's
equation of motion for the colliding particles reads
\begin{equation}
\frac{m}{2}\ddot{\xi}+F\left(\dot{\xi},\xi\right)=0\,;~~~~\xi(0)=0\;;~~~~ \dot{\xi}=v
\label{eq:EOM}
\end{equation}
where $v$ is the normal component of the initial relative velocity,
\begin{equation}
v\equiv \frac{\left[\vec{v}_1(0)-\vec{v}_2(0)\right]\cdot\left[\vec{r}_1(0)-\vec{r}_2(0)\right]}{2R}\,.
\label{eq:vdef}
\end{equation}
The coefficient of restitution follows from the solution of Eq. \eqref{eq:EOM}, 
\begin{equation}
\varepsilon= -\left.\dot{\xi}(t_c)\right/v\,,
\label{eq:COR1}
\end{equation}
where $t_c$ is the duration of the collision. 

Thus, the coefficient of restitution due to a specified interaction force law
can be determined by solving the equation of motion
\cite{RamirezEtAl:1999,SchaeferDippelWolf:1995,SchwagerPoeschel:1998}. In
general, $\varepsilon$ is a function of the material parameters, the particle
masses and radii and the impact velocity $v$.

\section{Linear dashpot model}

Consider the {\em linear
  dashpot} force,
\begin{equation}
    F\left(\xi,\dot{\xi}\right) =-k \xi - \gamma \dot{\xi}\,.
  \label{eq:lin_dashpot}
\end{equation}
This force is problematic as a particle interaction model since particles made
of a linear-elastic material do not reveal a linear repulsive force, neither
in tree dimensions \cite{Hertz:1882} nor in 2D \cite{Engel:1978}. The
subsequent analysis can be also performed for more realistic force models,
such as viscoelastic forces in 3d \cite{BrilliantovEtAl:1996} and 2d
\cite{Schwager} which will be published elsewhere.

The linear dashpot model is of interest here because this force leads to a
{\em constant} coefficient of restitution, e.g.
\cite{SchaeferDippelWolf:1995}, that is, $\varepsilon$ does not depend on the
impact velocity but only on the material parameters $k$ and $\gamma$. The
constant coefficient of restitution in turn is the preferred model in both the
Kinetic Theory of granular gases and also in event-driven simulations of
granular matter. In contrast, more realistic force models lead to an
impact-velocity dependent coefficient of restitution
\cite{LudingClementBlumenRajchenbachDuran:1994DISS,RamirezEtAl:1999,SchwagerPoeschel:1998,Taguchi:1992JDP,Tanaka:1983}.

The equation of motion corresponding to Eq. \eqref{eq:lin_dashpot}
\begin{equation}
  m \ddot{\xi} + 2\gamma\dot{\xi} + 2 k\xi = 0
\end{equation}
with initial conditions 
\begin{equation}
  \xi(0)  =0\,;~~~~~
  \dot{\xi}(0) = v  
\end{equation}
has two solutions. With the abbreviations 
\begin{equation}
  \omega_0^2\equiv\frac{2k}{m}\,;~~~~~\beta\equiv
  \frac{\gamma}{m}\,;~~~~\omega\equiv\sqrt{\omega_0^2-\beta^2}\,.  
\end{equation}
we can write the solution for the case of low damping ($\beta<\omega_0$) as
\begin{equation}
  \label{eq:dashpotSolutionlowdamping}
  \xi(t)=\frac{v}{\omega}e^{-\beta t} \sin\omega t\,,
\end{equation}
while for the case of high damping ($\beta>\omega_0$) we have 
\begin{equation}
  \label{eq:dashpotSolutionhighdamping}
  \xi(t)=\frac{v}{\Omega}e^{-\beta t} \sin\Omega t\,,
\end{equation}
with
\begin{equation}
  \Omega = \sqrt{\beta^2-\omega_0^2}\,.
\end{equation}
To determine the coefficient of restitution according to Eq. \eqref{eq:COR1}
we need the duration of the collision $t_c$. A natural choice seems to be
\begin{equation}
  \xi\left(t_c^0\right)=0\,;~~~~~ t_c^0>0 \,,
  \label{eq:dashpotEndOfColl}
\end{equation}
that is, the collision is finished when
$\left|\vec{r}_i-\vec{r}_j\right|=2R$.  With this condition we obtain
$t_c^0=\pi/\omega$ and the coefficient of restitution
\begin{equation}
  \label{eq:dashpotCOR}
  \varepsilon_{\rm d}^0\equiv\frac{\dot{\xi}\left(t_c^0\right)}{\dot{\xi}(0)}=e^{-\beta\pi/\omega}\,,
\end{equation}
for the case of low damping which is widely used, e.g. in
\cite{SchaeferDippelWolf:1995} and many others.

\begin{figure*}[t]
\begin{center}
\begin{tabular}{
    p{0.1\textwidth}
    p{0.1\textwidth}
    p{0.1\textwidth}
    p{0.1\textwidth}
    p{0.1\textwidth}
    p{0.1\textwidth}
    p{0.1\textwidth}
    }
&&&&&&\\[-18pt]
\centering\includegraphics[height=1.3cm,angle=270]{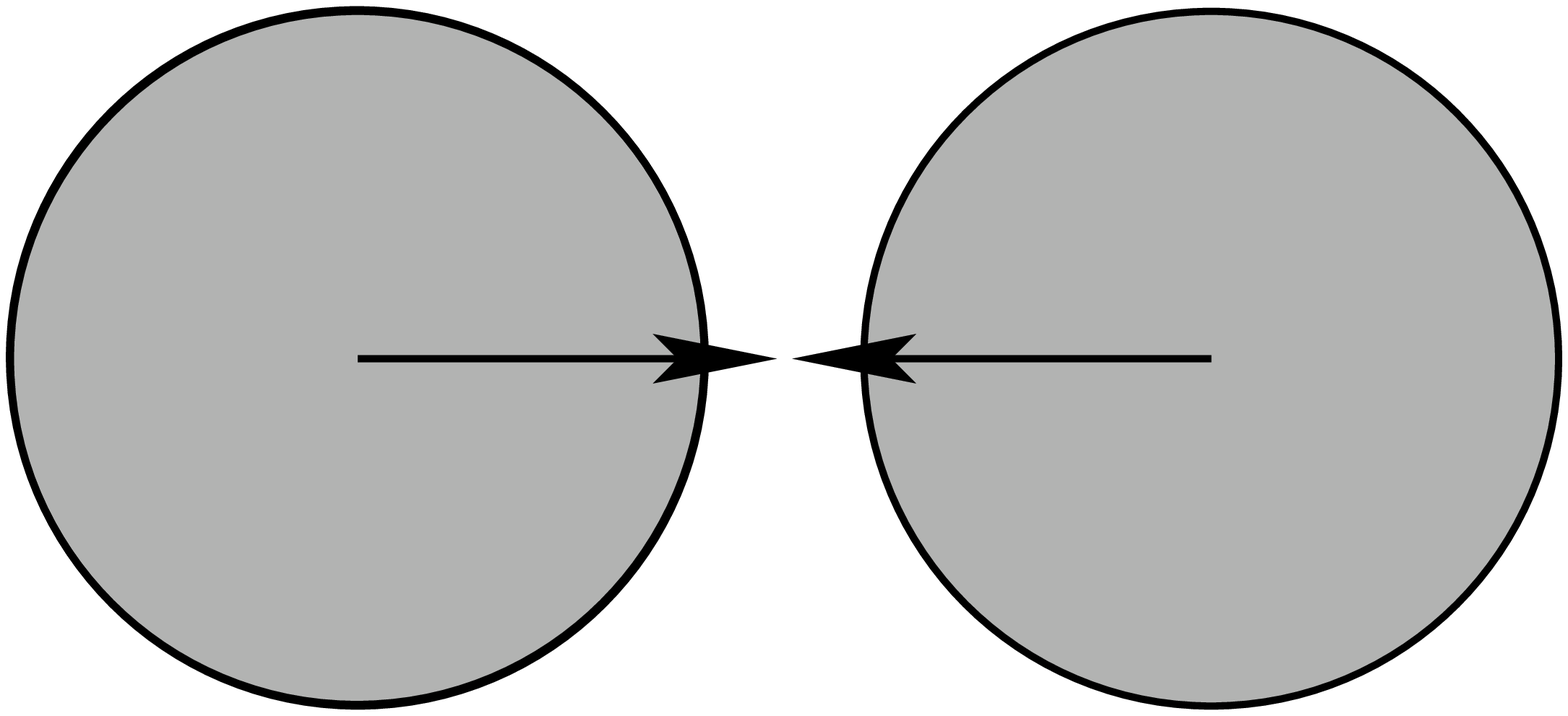} &
\centering\includegraphics[height=1.3cm,angle=270]{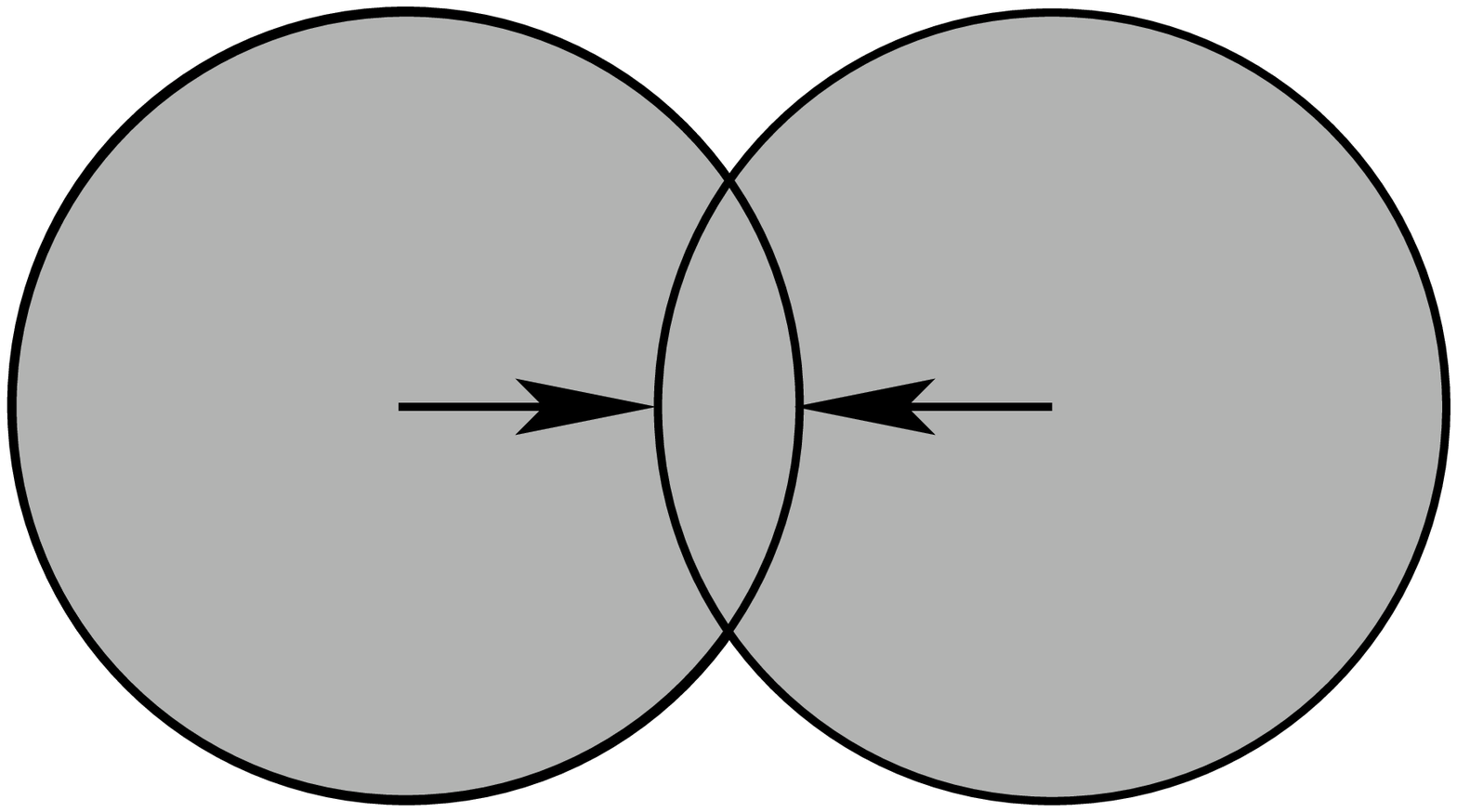} &
\centering\includegraphics[height=1.3cm,angle=270]{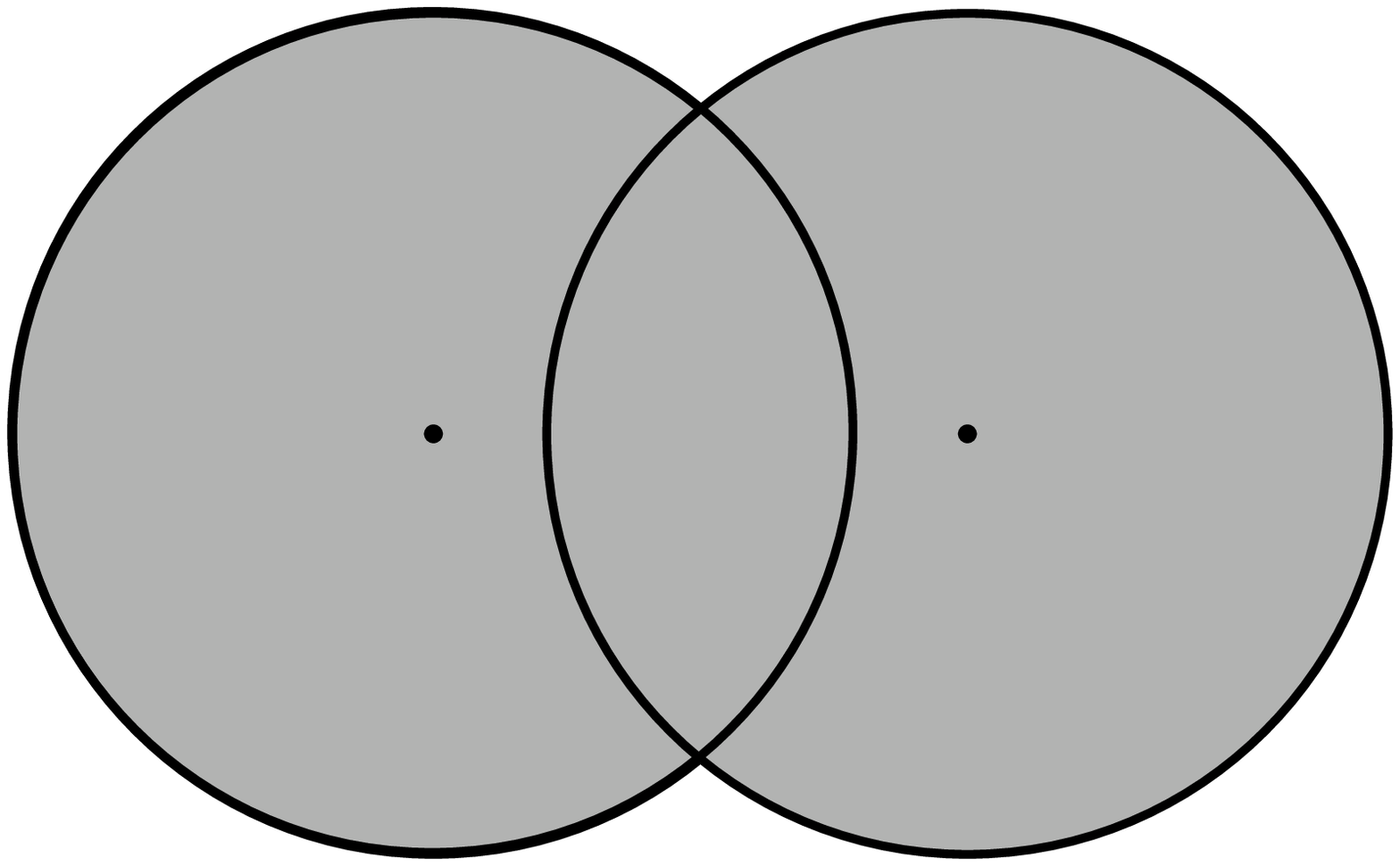} &
\centering\includegraphics[height=1.3cm,angle=270]{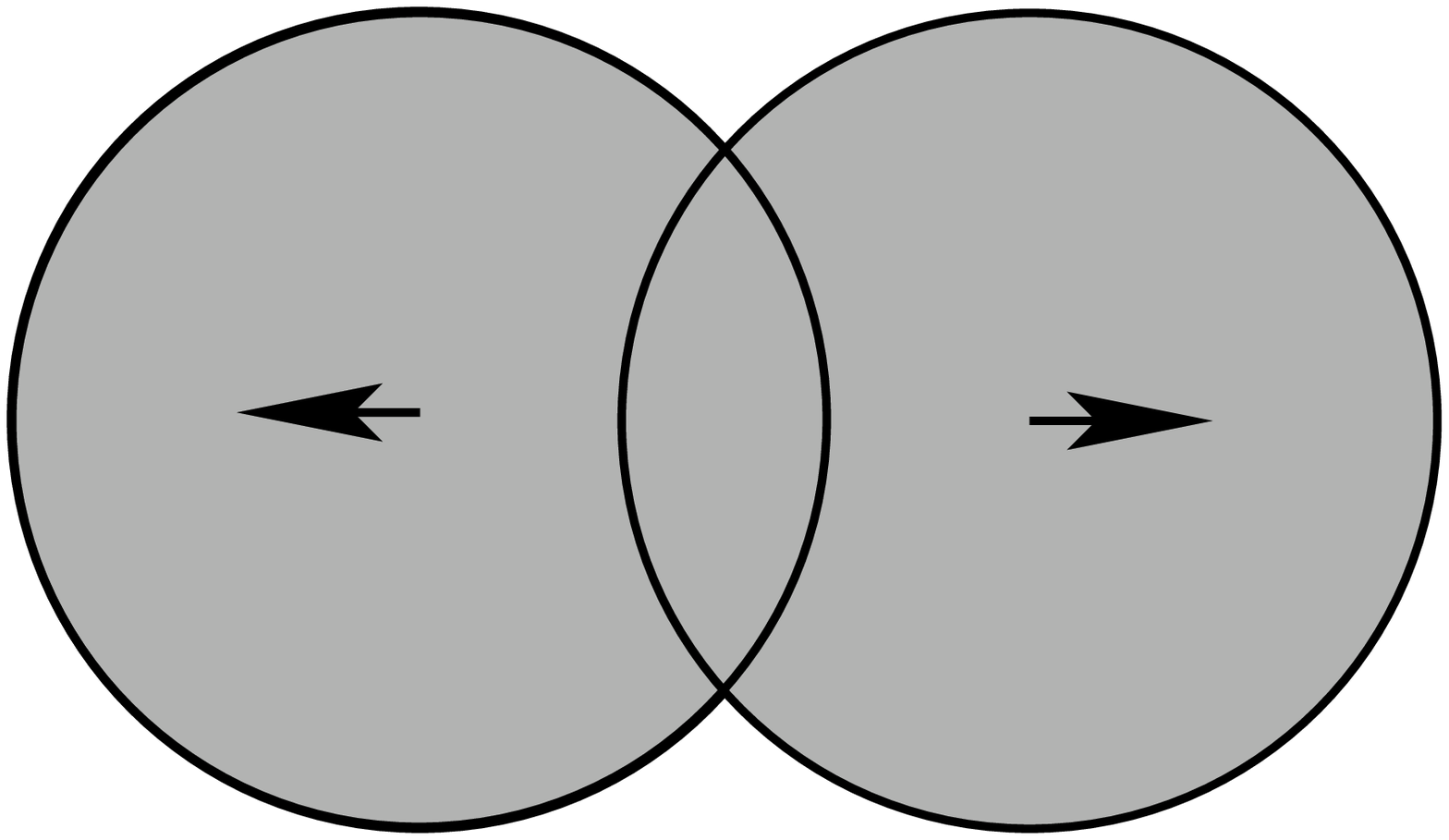} &
\centering\includegraphics[height=1.3cm,angle=270]{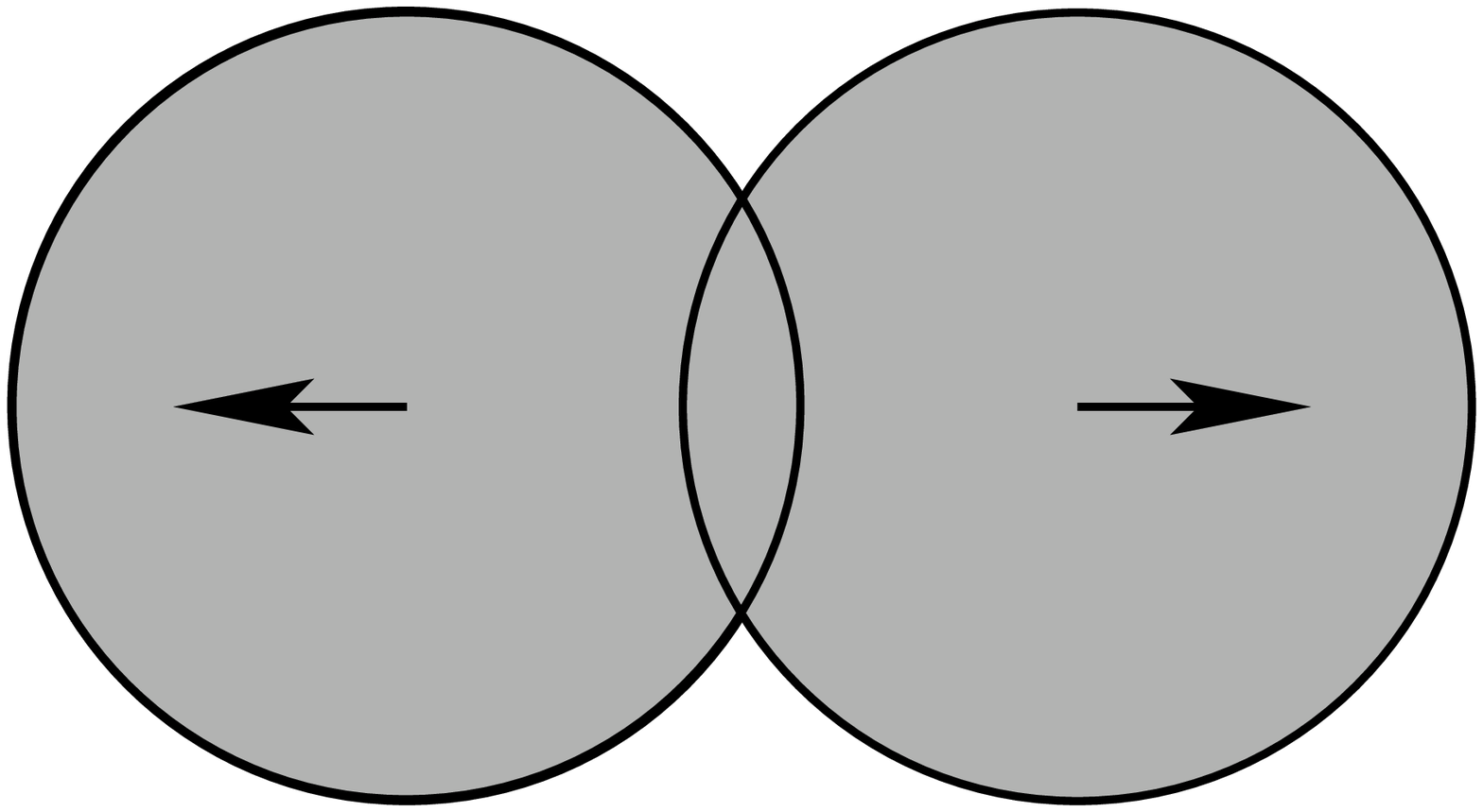} &
\centering\includegraphics[height=1.3cm,angle=270]{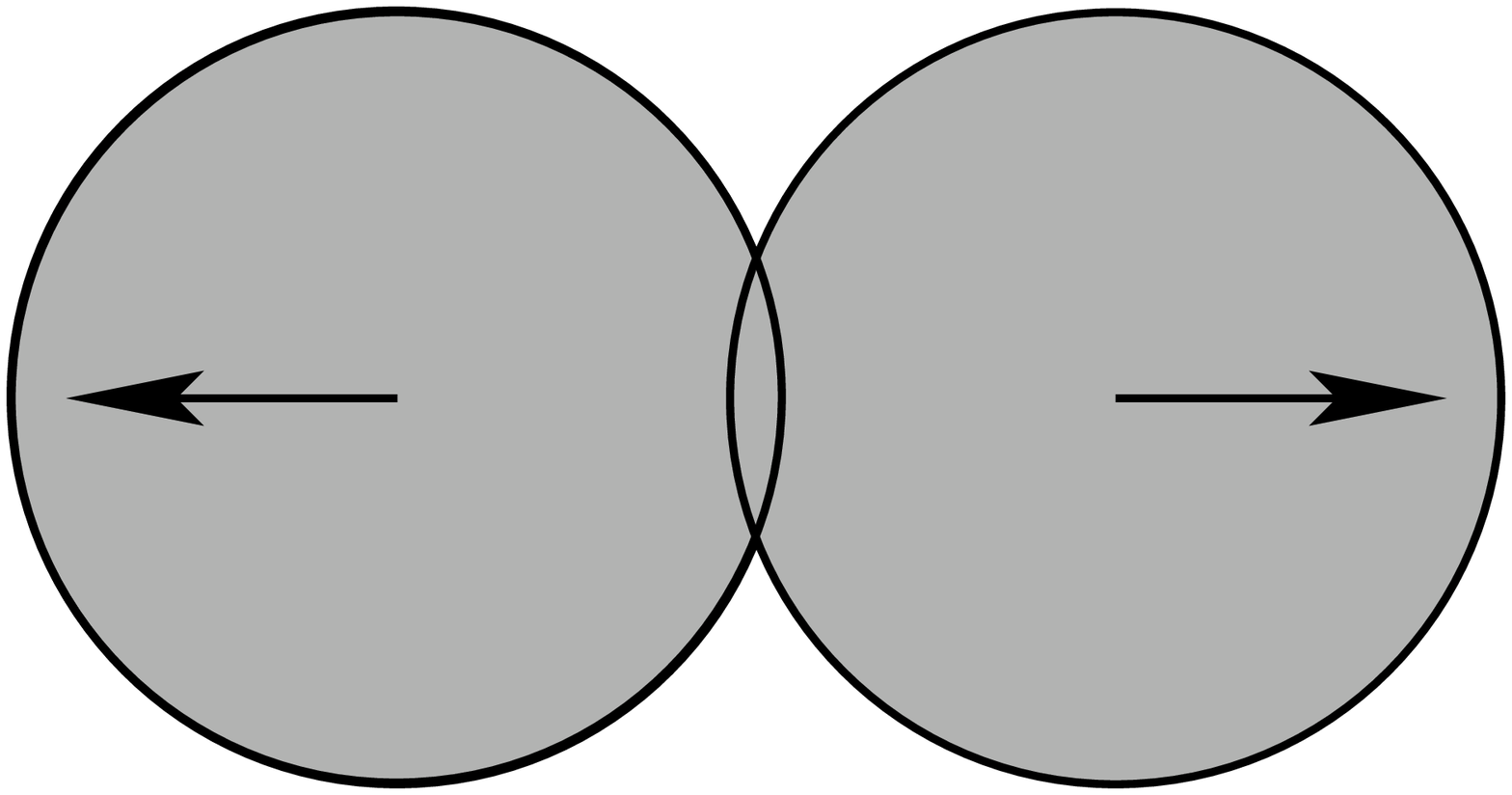} &
\centerline{\includegraphics[height=1.3cm,angle=270]{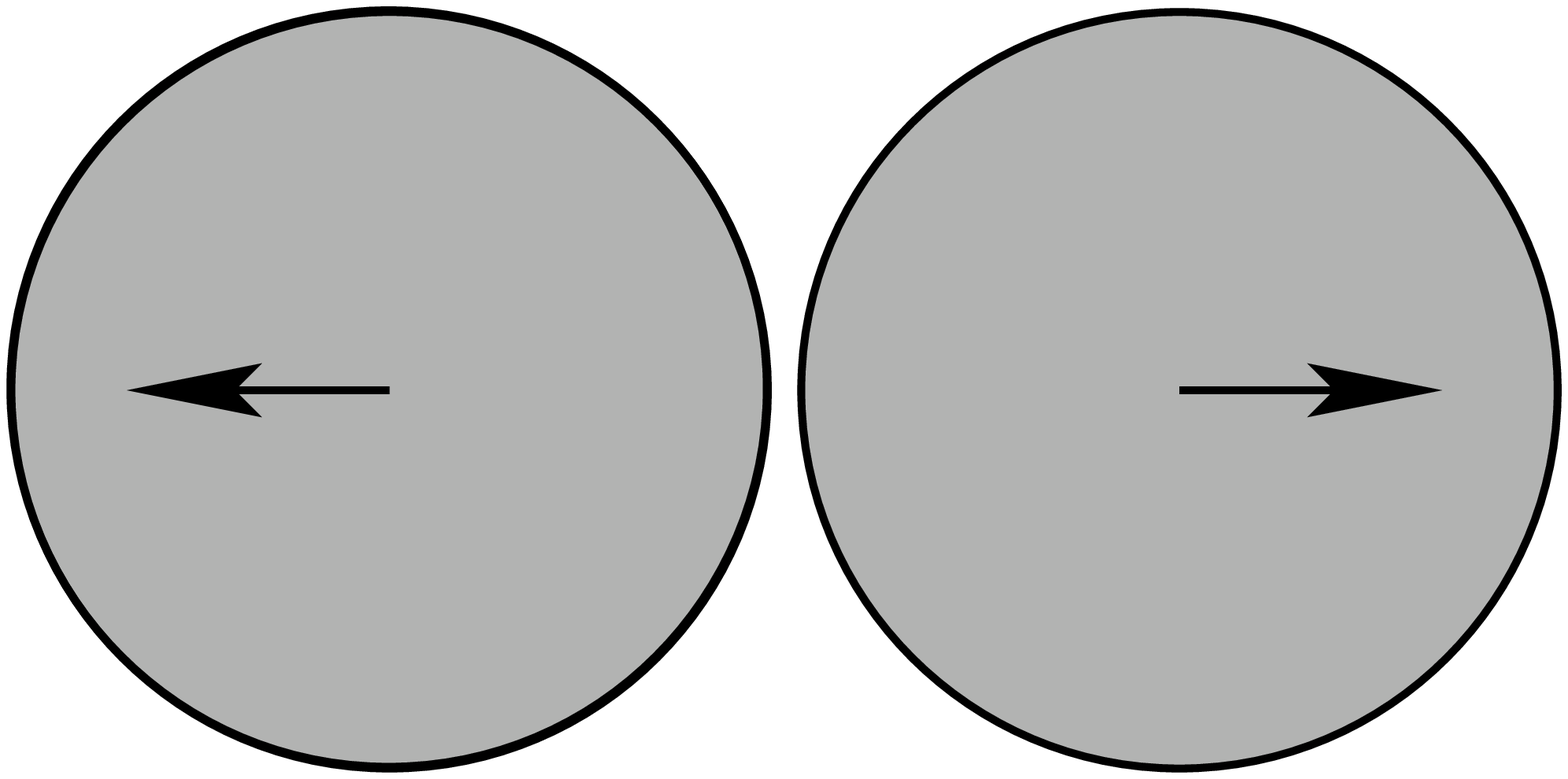}}\\
\centerline{$\xi<0$} &
\centerline{$\xi>0$} &
\centerline{$\xi^{\rm max}>0$} &
\centerline{$\xi>0$} &
\centerline{$\xi>0$} &
\centerline{$\xi>0$} &
\centerline{$\xi<0$} \\
\centerline{$\dot{\xi}=g>0$} &
\centerline{$\dot{\xi}>0$} &
\centerline{$\dot{\xi}=0$} &
\centerline{$\dot{\xi}<0$} &
\centerline{$\dot{\xi}^*<0$} &
\centerline{$\displaystyle\dot{\xi}>\dot{\xi}^*\atop\displaystyle\dot{\xi}<0~$} &
\centerline{$\dot{\xi}<0$}\\
\centerline{$F=0$} &
\centerline{$F>0$} &
\centerline{$F>0$} &
\centerline{$F>0$} &
\centerline{$F=0$} &
\centerline{\fbox{$F<0$}} &
\centerline{$F=0$}\\
\centering\includegraphics[height=1.3cm,angle=270]{figs/bug1.eps} &
\centering\includegraphics[height=1.3cm,angle=270]{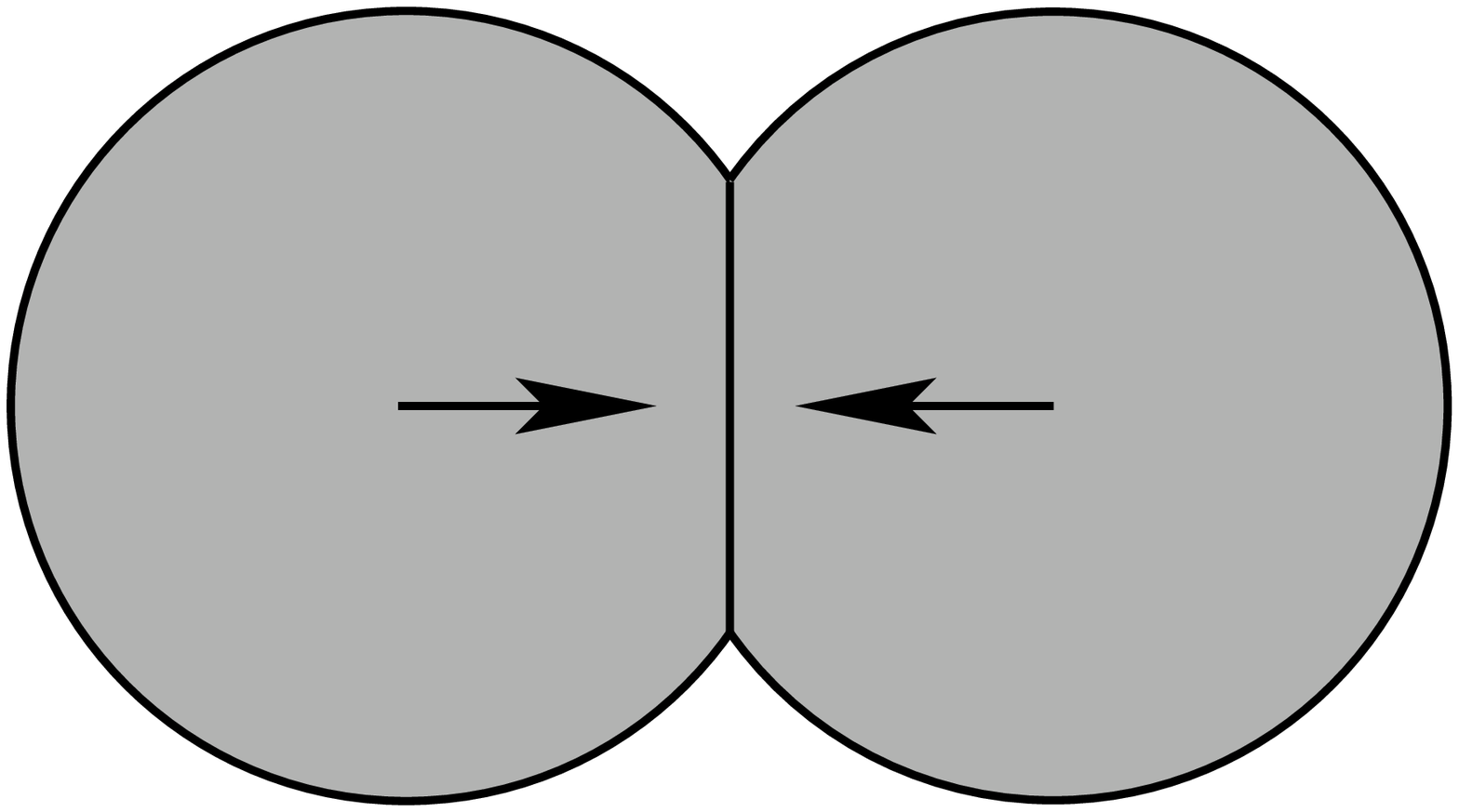} &
\centering\includegraphics[height=1.3cm,angle=270]{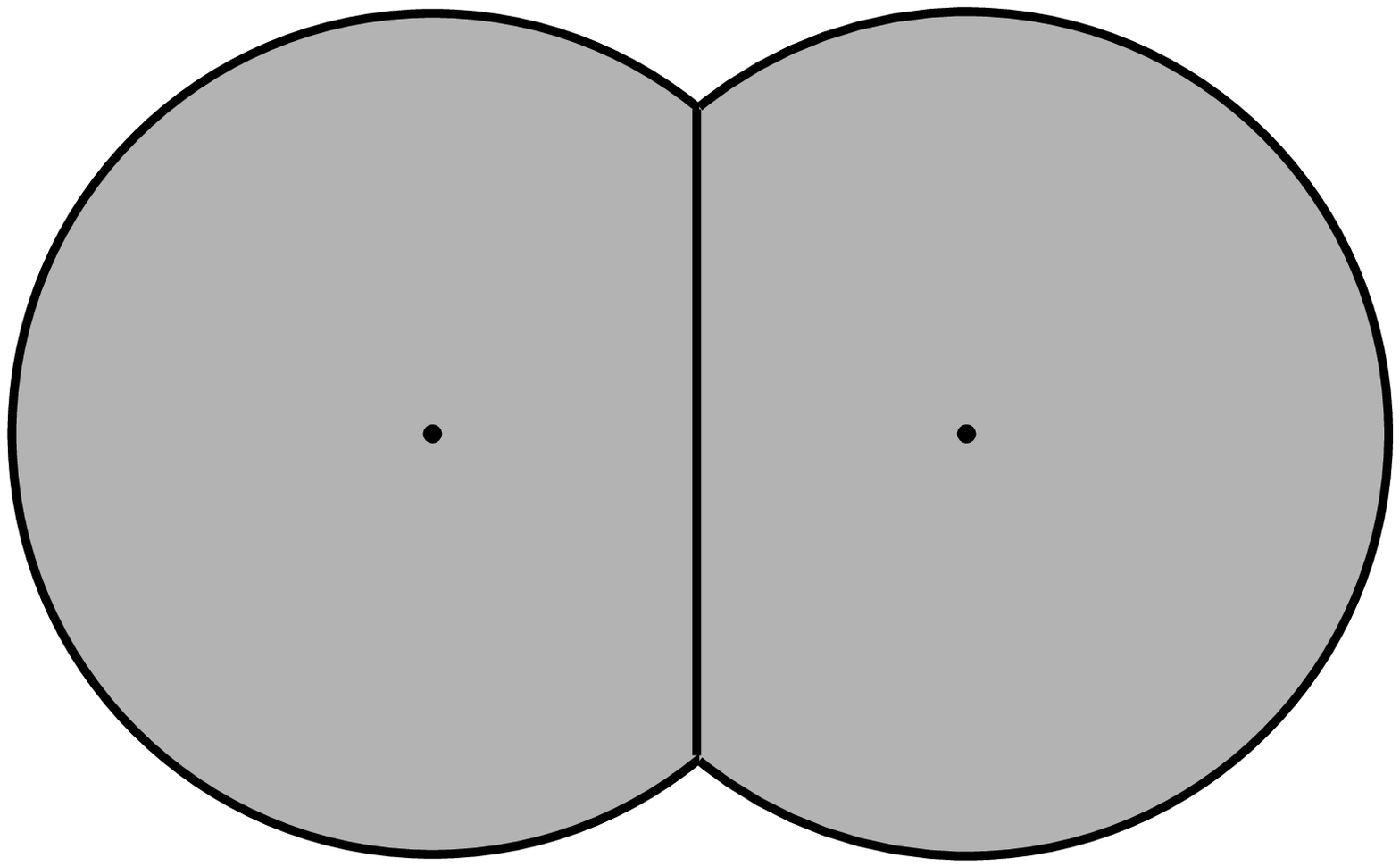} &
\centering\includegraphics[height=1.3cm,angle=270]{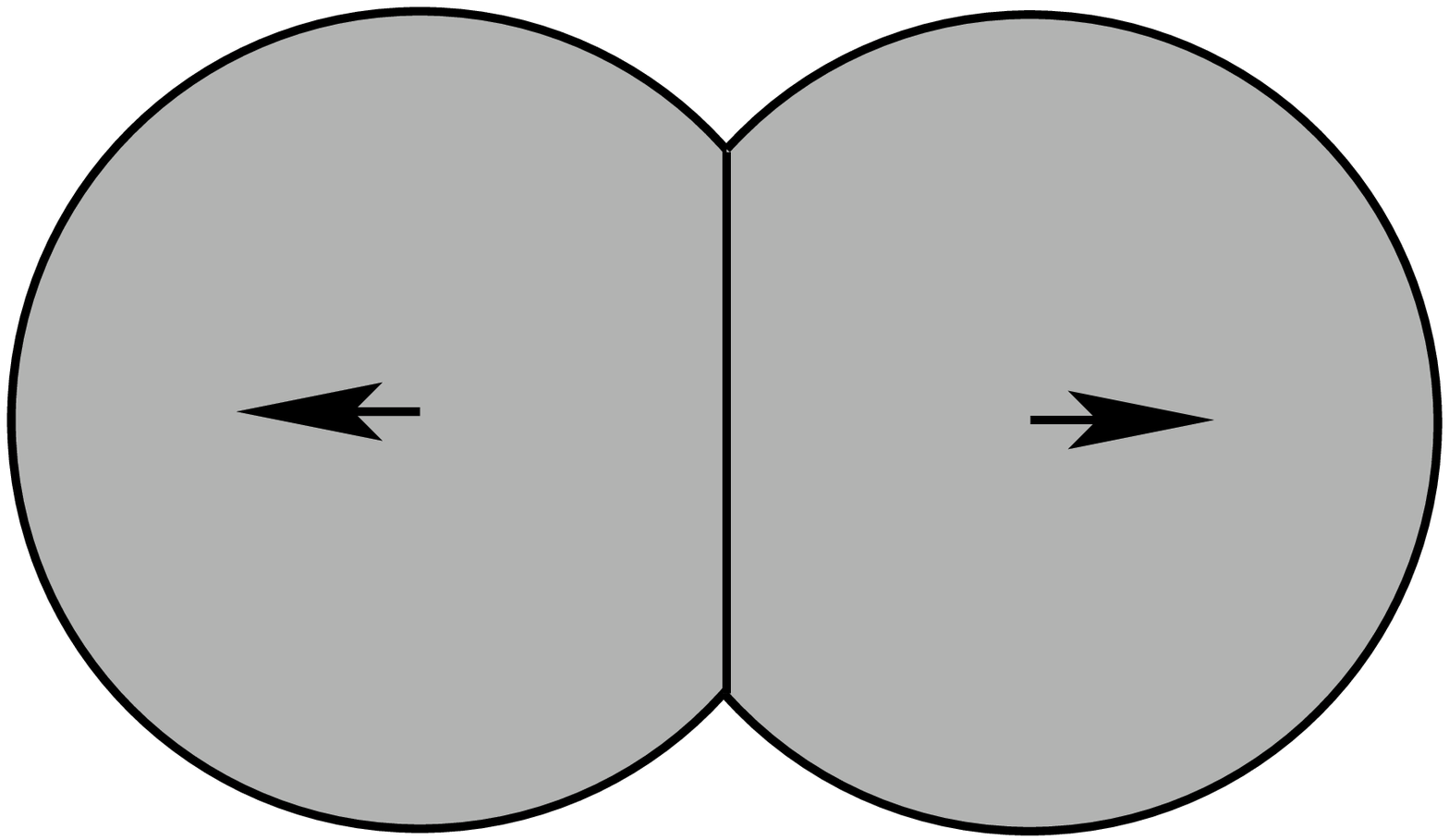} &
\centering\includegraphics[height=1.3cm,angle=270]{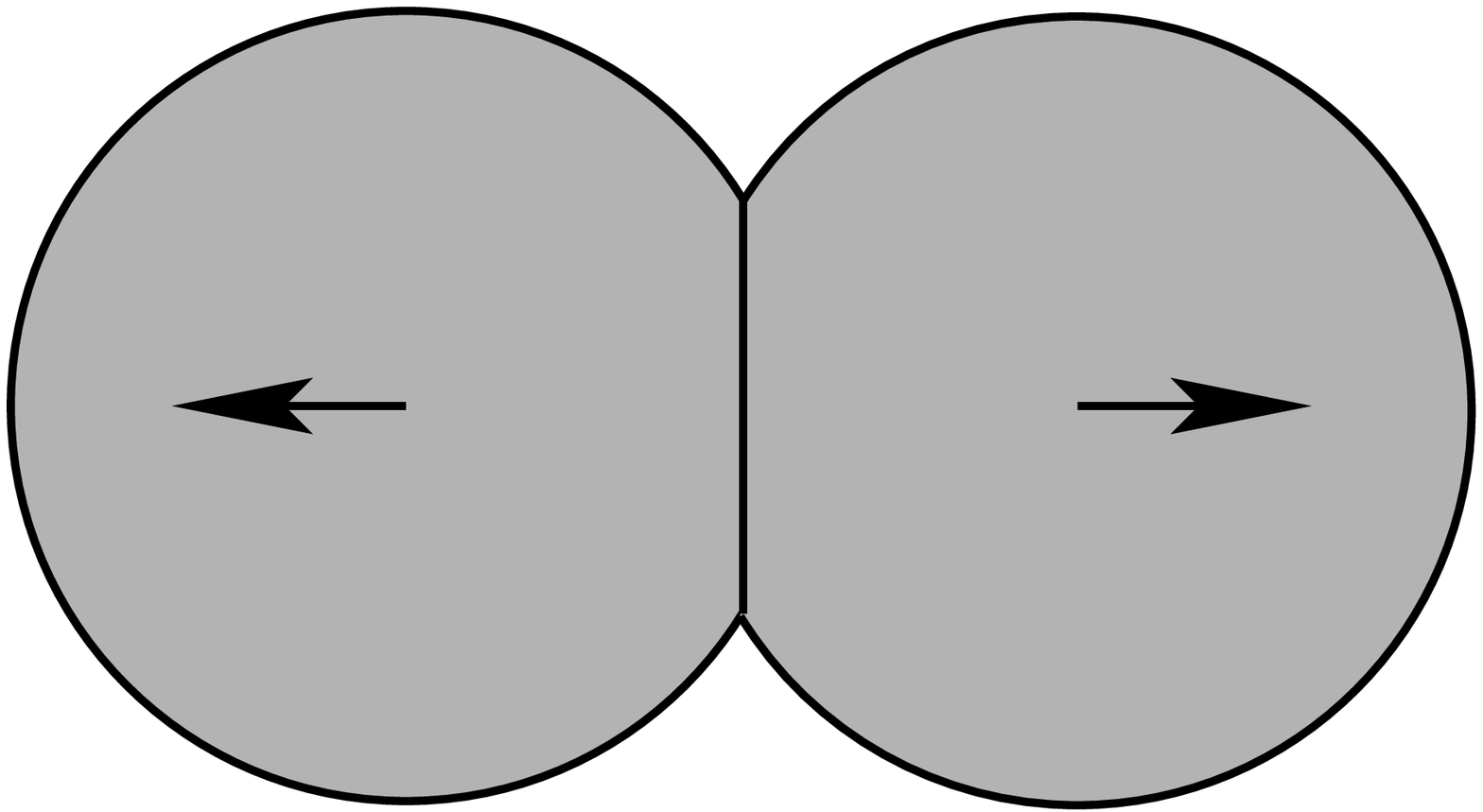} &
\centering\includegraphics[height=1.3cm,angle=270]{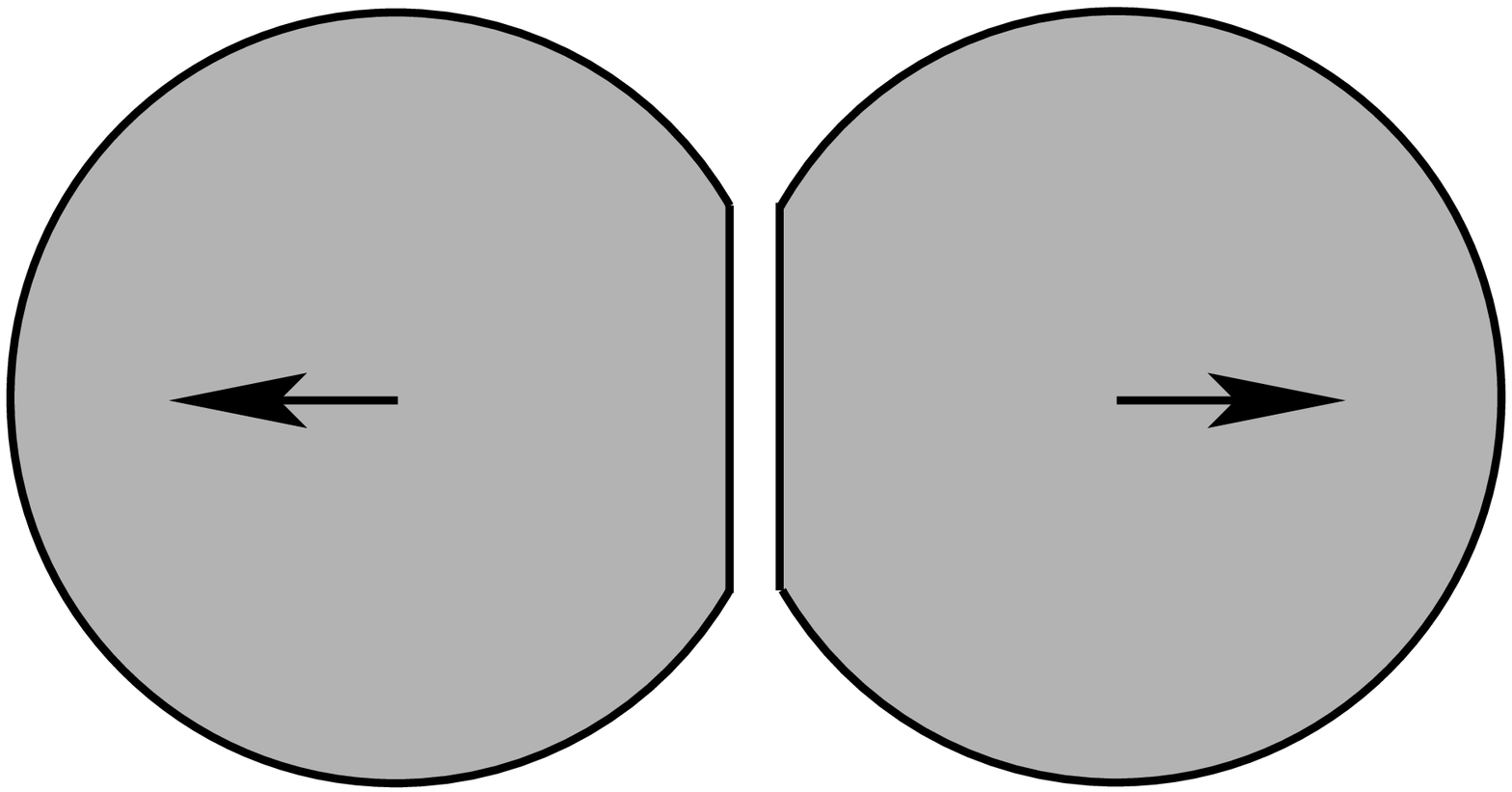} &
\centerline{\includegraphics[height=1.3cm,angle=270]{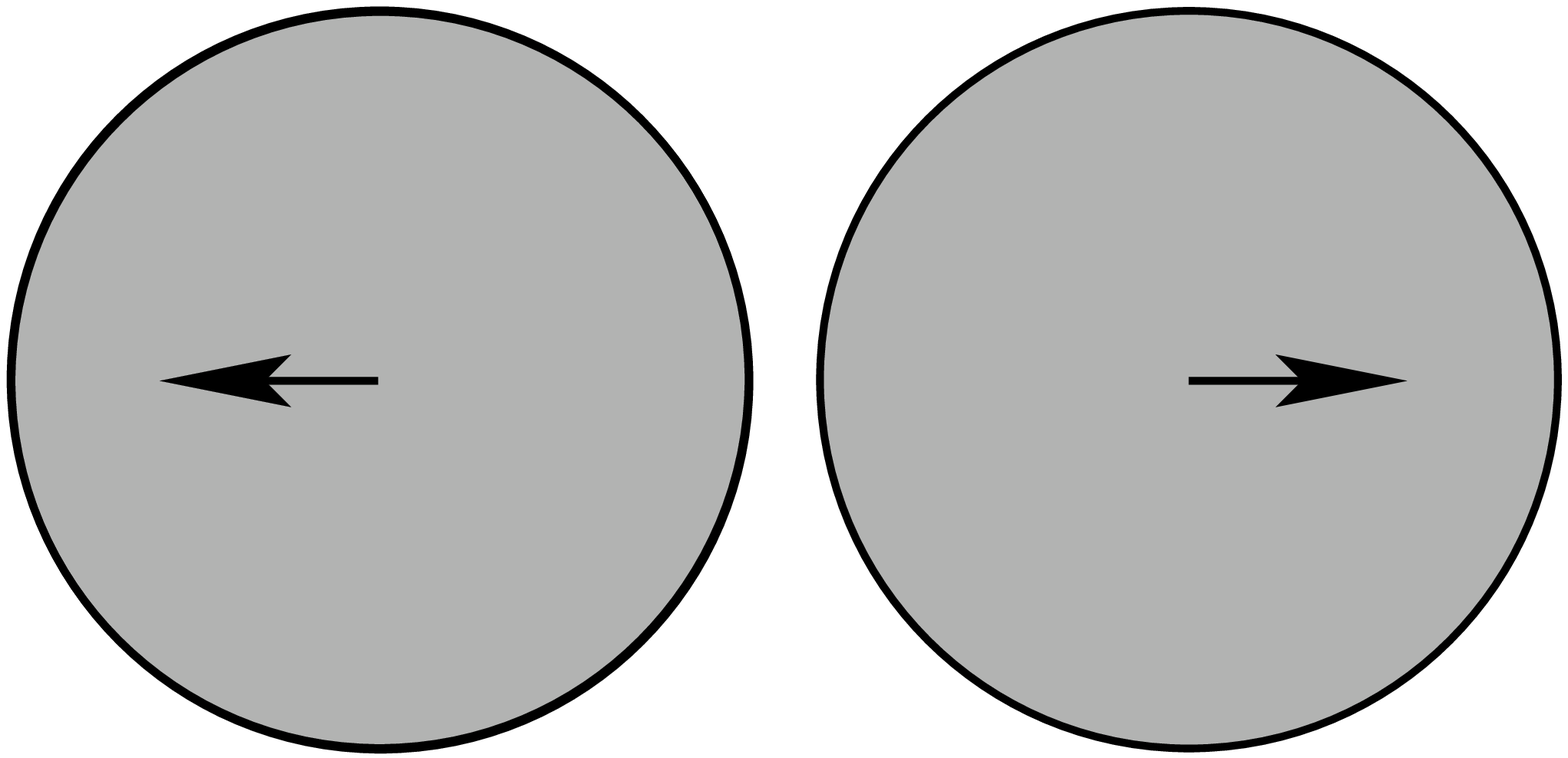}}\\
\centerline{$F=0$} &
\centerline{$F>0$} &
\centerline{$F>0$} &
\centerline{$F>0$} &
\centerline{$F=0$} &
\centerline{\fbox{$F=0$}} &
\centerline{$F=0$}\\[-1pc]
\end{tabular}
\end{center}  
  \caption{Sketch of a particle collision. The upper part of the figure visualizes the condition \eqref{eq:dashpotEndOfColl} for the end of the collision. As an artifact of the model, there appear attracting forces towards the end of the collision (boxed notation).  Lower part: The condition \eqref{eq:dashpotEndOfColl1} for the end of the collision avoids this artifact. 
}
  \label{fig:Fnbug}
\end{figure*}

\begin{figure}
\centerline{\includegraphics[width=0.9\columnwidth,clip=]{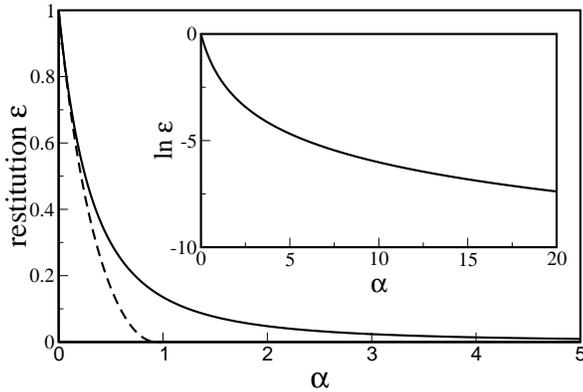}
}
\caption{Coefficient of restitution as function of $\alpha=\beta/\omega_0$. The dashed curve is the result for the wrong termination condition $\xi(t_c^0)=0$. Obviously, for high damping $\alpha>1$ we have dissipative capture. The inlay shows $ln\varepsilon$ as a function of $\alpha$ for the correct termination condition. There is no dissipative capture even for very high damping.}
\label{fig:eps}
\end{figure}

Unfortunately, condition \eqref{eq:dashpotEndOfColl} cannot be correct since
at time $t_c^0=\pi/\omega$ the interaction force is negative,
\begin{equation}
  \label{eq:dashpotAcc}
  m\ddot{\xi}\left(\frac{\pi}{\omega}\right)=-\frac{2mv\beta}{\omega}e^{-\beta\pi/\omega} <0\,,
\end{equation}
see Fig. \ref{fig:Fnbug}, which contradicts the assumption of exclusively
repulsive interaction between granular particles. Moreover, for the case of
high damping Eq. \eqref{eq:dashpotEndOfColl} has no real solution, that is,
the particles stick together after a head-on collision (see Fig.
\ref{fig:eps}). One can call this situation {\em dissipative capture}.

To avoid these artifacts, instead of Eq. \eqref{eq:dashpotEndOfColl}, we now
explicitely take  into account the condition of no attraction. That means,
instead of Eq. \eqref{eq:lin_dashpot} we have to use the force
$F^*=\max(0,~F)$, with $F$ given by Eq. \eqref{eq:lin_dashpot}. This
expression as is in common use in (force-based) Molecular Dynamics
simulations, assures that only repulsive forces act. It is clear that once the
force becomes zero, the collision is finished,
\begin{equation}
  \label{eq:dashpotEndOfColl1}
  \ddot{\xi}(t_c)=0\,;~~~~~t_c>0\,;~~~~~\dot{\xi}(t_c)<0\,,
\end{equation}
that is, the collision is complete when the repulsive force between the
particles vanishes and the surfaces of the particles separate from one
another. This condition takes, thus, into account that Eq. \eqref{eq:lin_dashpot}
applies to {\em particles in contact}.

For the case of low damping Eqs. \eqref{eq:dashpotSolutionlowdamping} and \eqref{eq:dashpotEndOfColl1} yield
\begin{equation}
  \label{eq:indirectdashpotDuration}
  \tan\omega t_c=-\frac{2\beta\omega}{\omega^2-\beta^2}~.
\end{equation}
One has to take care to select the correct branch of the arc tangent for $\beta<\omega_0/\sqrt{2}$ (or $\beta<\omega$) and for $\beta>\omega_0/\sqrt{2}$. The solution reads
\begin{equation}
  \label{eq:dashpotDuration}
  t_c = \begin{cases}
    \displaystyle\frac{1}{\omega}\left(\pi - \arctan\displaystyle\frac{2\beta\omega}{\omega^2-\beta^2}\right) & \mbox{for~~~} \displaystyle\beta<\frac{\omega_0}{\sqrt{2}} \\[0.5cm]
      \displaystyle\frac{1}{\omega}\arctan\displaystyle\frac{2\beta\omega}{\omega^2-\beta^2} & \mbox{for~~~} \displaystyle\beta>\frac{\omega_0}{\sqrt{2}}
\end{cases}
\end{equation}
For the coefficient of restitution we find
\begin{equation}
  \varepsilon_d = e^{-\beta t_c}
\end{equation}
regardless of the branch of the solution of Eq.
\eqref{eq:indirectdashpotDuration}. Together with the solution for the case of
high damping which can be obtained by straight-forward computations we obtain
the coefficient of restitution:
\begin{equation}
  \varepsilon_d = \begin{cases}
    \displaystyle\exp\left[-\frac{\beta}{\omega}\left(\pi - \arctan\displaystyle\frac{2\beta\omega}{\omega^2-\beta^2}\right)\right] & \!\!\!\!\!\!\mbox{for~}  \displaystyle\beta<\frac{\omega_0}{\sqrt{2}} \\[0.3cm]
    \displaystyle\exp\left[-\frac{\beta}{\omega}\arctan\displaystyle\frac{2\beta\omega}{\omega^2-\beta^2}\right]
    & \!\!\!\!\!\!\mbox{for~} \displaystyle \beta\in\left[\frac{\omega_0}{\sqrt{2}},\omega_0\right]
\\[0.3cm]
    \displaystyle\exp\left[-\frac{\beta}{\Omega}\ln\frac{\beta+\Omega}{\beta-\Omega}\right] & \!\!\!\!\!\!\mbox{for~} \beta>\omega_0
    \end{cases}
  \label{eq:dashpotCOR1}
\end{equation}
In accordance with physical intuition we have $\varepsilon_d >
\varepsilon_d^0$ since the result given in Eq. \eqref{eq:dashpotCOR1} does not
suffer from the artifact of unphysical attractive forces implied in Eq.
\eqref{eq:dashpotCOR}. The unphysical dissipative capture does not occur --
even for large damping constant the particles separate eventually. For very
high damping the coefficient of restitution can be approximated as
\begin{equation}
  \label{eq:asymptotics}
  \varepsilon_d \approx \frac{\omega_0^2}{4\beta^2}\mbox{~~~~~for~~~} \beta\gg\omega_0
\end{equation}

Fig. \ref{fig:eps} shows both the (wrong) coefficient of restitution as
derived from criterion \eqref{eq:dashpotEndOfColl} (dashed line) and the
coefficient of restitution derived from the correct criterion
\eqref{eq:dashpotEndOfColl1} (full line). As the solution Eq.
\eqref{eq:dashpotCOR1} only depends on the ratio $\alpha =
\beta/\omega_0$ we draw the coefficient of restitution as a function of this
parameter. Figure \ref{fig:eps} shows that the unphysical attraction indeed
yields a too low coefficient of restitution. The inset shows a logarithmic
plot of the same curve showing that there is no capture if the correct
criterion is applied. 

\section{Conclusion}

Assuming a linear dashpot interaction force we computed $\varepsilon$ as a function of the elastic and dissipative material properties by integrating Newton's equation of motion for a pair of colliding particles. The condition for the duration $t_c$ of the collision $\xi\left(t_c^0\right)=0$, that is, the interaction stops when the particles have the distance $\left|\vec{r}_i-\vec{r}_j\right|=2R$, is widely used in the literature, however, it leads to erroneous attracting forces between the particles and, thus, to wrong values for the coefficient of restitution. In extreme cases, the result may become even qualitatively incorrect since despite the interaction force being purely repulsive, the particles may agglomerate. In this state one would need to supply energy to separate the particles from one another. 

To account for the fact that the interaction force can never become attractive, we use the condition  $\ddot{\xi}(t_c)=0$ which avoids the mentioned artifacts. This condition takes into account that the collision may be completed even before $\xi=0$, that is, the surfaces of the particles lose contact slightly before the distance of their centers exceeds the sum of their radii. Thus, the deformation of the particles may last longer than the time of contact and the particles gradually recover their spherical shape only {\em after} they lost contact.

The resulting coefficient of restitution is always larger than the value based
on $\xi\left(t_c^0\right)=0$. In agreement with the repulsive character of the
interaction force, the coefficient of restitution is always well defined and non-zero.

\bibliographystyle{spbasic}
\bibliography{epsilon_dashpot}   

\end{document}